\documentstyle[12pt,aaspp4]{article}

\begin{document}

\title{The Observed Size and Shape of GRB Afterglow}
\author{Re'em Sari}
\affil{Racah Institute, Hebrew University, Jerusalem 91904, Israel \\
and\\
Harvard-Smithsonian Center for Astrophysics, 
60 Garden Street, Cambridge MA 02138, U.S.A.
}
\authoremail{sari@shemesh.fiz.huji.ac.il}

\begin{abstract}
  The detection of delayed emission in X-ray, optical and radio wave
  length, ``afterglow'', following a $\gamma$-ray burst can be
  described by the emission of a relativistic shell decelerating upon
  collision with the ISM. We show that the observed radiation surface
  have well defined bright edges. We derive an explicit expression for
  the size as a function of time, and obtain the surface brightness
  distribution. This might be directly observed if the burst occurs at
  small redshift so that its radio signal can be resolved. The size
  and shape are relevant for detailed analysis of scintillation or
  microlensing. We show that the effective Lorentz factor depends on
  the observed frequency and it is higher for frequencies above the
  synchrotron typical frequency (optical and X-ray) than for low
  frequencies (radio). Consequently transition to non relativistic
  evolution, will be observed first in low frequencies and only a
  factor of $\sim 2$ later in the high frequencies.
\end{abstract}


\section{Introduction}
The detection of delayed emission in X-ray optical and radio wave
length, ``afterglow'', following a $\gamma$-ray burst is reasonably
described by the emission of a relativistic shell decelerating upon a
collision with the ISM (Waxman 1997a,b, Wijers Rees and M\'esz\'aros
1997, Katz and Piran 1997). The radio observations show substantial
oscillations of order unity over the first few weeks (Frail et. al.
1997), while the optical measurements at the same time follow a smooth
regular decline.  It is therefor likely that the nature of these
oscillations is not intrinsic to the source. These oscillations might
be explained by scintillation in our galaxy. Goodman's estimates
(1997) show that the size of the observed emitter is in the limit
between being a large object, subject only to refractive
scintillation, and being a small one, subject to diffractive
scintillation that can account for fluctuations of order unity. In
order to correctly account for this scintillation a good estimate for
the size of the observed object is needed. In addition, the size of
the object is relevant for estimates of the observed luminosity if the
emission is self absorbed (Katz \& Piran 1997). The size is also
important in concern with microlensing (Loeb \& Perna 1997). In
section 2 we show that the observed radiation has well defined
boundaries and derive an explicit expression for its size. In section
3 we discuss the brightness distribution within this limited size.  We
find the flux averaged radius and Lorentz factor of the material
emitting the radiation. We show that this effective emission radius
for observed frequencies above the peak frequency is smaller than for
observed frequencies below the peak frequency.

\section{The Size of The Afterglow}
We consider first the well know result (Rees 1966) of the observed
size of a thin shell with constant velocity $\beta c$. This relation
is simple to derive and it gives the basic picture. It is also
appropriate for non relativistic velocities. After a time $t$ the
shell is located at radius $R=\beta c t$. A photon that is emitted at
an angle $\theta$ from the line of sight (connecting the observer and
the source) will reach the observer at time
\begin {equation}
\label{ellipse}
T=t-R\mu/c =R(1-\beta\mu)/(\beta c),
\end{equation}
where $\mu=\cos \theta$. This expression describes the locus of points
from which the radiation reaches the observer at the same time (see
upper frame of Figure 1). It is an ellipse whose long axis is along
the line of sight. The source is located at the far focal point.  The
observed size of the emitting area is the small axis of the ellipse
$\beta\gamma c T$ .

\begin{figure}
\begin{center}
\plotone{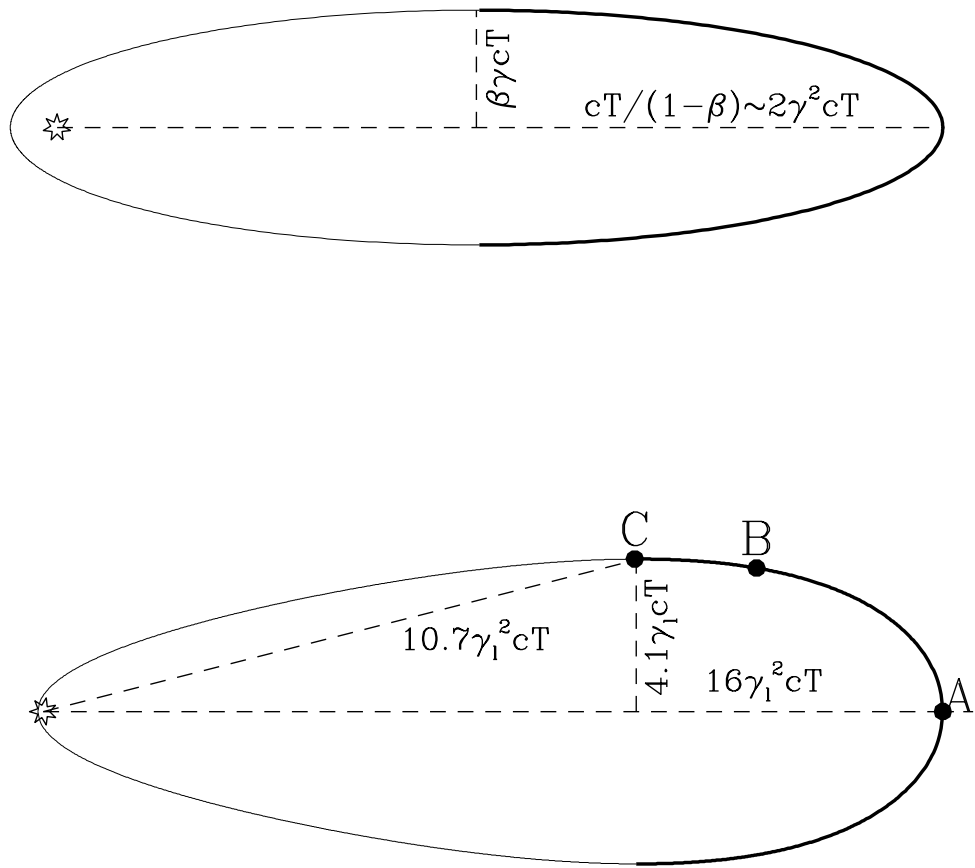}
\caption{The locus of points from which radiation from a thin shell arrives
  at the same time to the observer. The upper frame is for a shell
  with constant speed and the lower frame is for a decelerating shell.
  The observer is located far on the right side. The sizes for the
  decelerating case are under the approximation of relativistic
  motion. $\gamma_l$ is the Lorentz factor of the emitting material on
  the line of sight,i.e, at the point A. Point B is the average
  emission point for low frequencies while point C is the average
  emission point for high frequencies.}
\end{center}
\end{figure}
In the afterglow scenario a shell accumulates ISM mass and decelerates
with $\gamma\propto R^{-3/2}$.  Due to the accumulation of mass, the
Lorentz factor of the shock front is higher than the Lorentz factor of
the material behind it $\gamma$ by a factor of $\sqrt 2$.  As we have
recently pointed out (Sari 1997), the relation between the observed
time of arrival of photon from the line of sight and the radius at
which it was emitted is $R_l=16\gamma_l^2 c T$, where $\gamma_l$ is
the Lorentz factor of the material immediately behind the shock. The
subscript $l$ indicates that this relation is valid only for photons
arriving from the line of sight. An equivalent expression relating $R$
and the time of emittance $t$ is
\begin{equation}
\label{Roft}
R=c t/\left( 1+\frac 1 {16\gamma^2} \right).
\end{equation}
The arrival time $T$ of a photon emitted at an angle $\theta$ is given by
\begin {equation}
\label{Toft}
T=t-R\mu/c.
\end{equation}
Solving equation \ref{Roft} and \ref{Toft} we have
\begin{equation}
R(T,\mu)=\frac   {c T}    { 1-\mu +1/(16\gamma^2) }
\end{equation}
Substituting $\gamma$=$\gamma_l(R/R_l)^{-3/2}$, the above equation yields
\begin{equation}
\label{muofR}
1-\mu=\frac 1 {16 \gamma_l^2} \left[
\left( \frac R {R_l} \right)^{-1} -
\left( \frac R {R_l} \right)^{ 3}
\right].
\end{equation}
Equation \ref{muofR} is the decelerating analogy of the ellipse given in
equation  \ref{ellipse} for a constant velocity shell.  It describes an
elongated ``egg-shape'' shown in the lower frame of Figure 1.

The relevant size in the context of scintillation or microlensing
would be the distance from the line of sight $R_\perp$:
\begin{equation}
\label{Rpofmu}
R_\perp=R\sqrt{1-\mu^2} \cong 
\frac {\sqrt 2 R_l} {4 \gamma_l} \sqrt {
\left( \frac R {R_l} \right)   -
\left( \frac R {R_l} \right)^5
}
\end{equation}
The maximal value of the expression is obtained at
\begin{equation}
\frac R {R_l} =\left( \frac 1 5 \right) ^{1/4} \cong 0.67,
\end{equation}
where
\begin{equation}
R_{\perp , \max} = 4.1 \gamma_l c T
\end{equation}
This should be compared with the constant velocity shell 
$R_\perp,\max=\beta \gamma c T$.
A substitution of the expression for
$\gamma_l$ (Sari 1997) obtained by using the self similar solution of 
Blandford \& McKee (1976) yields:
\begin{equation}
\label{rperpmax}
R_{\perp , \max}=3.2 \times 10^{13} E_{52}^{1/8} n_1^{-1/8} (T/1{\rm s})^{5/8}
\end{equation}
This result is quite robust as it depend very weakly on the energy in
the system $10^{52}E_{52}$ erg and the ISM particle density $n_1 {\rm\ 
  cm^{-3}}$.  The only assumptions used are adiabatic evolution
described by the Blandford and McKee self similar solution, and the
assumption of high Lorentz factor.

\section{The Surface Brightness Distribution}

Equation \ref{rperpmax} gives the maximal $R_\perp$ from which radiation
will reach the observer at a given observer time.  The distribution of
the surface brightness within this radius requires additional
assumptions concerning the emitted spectrum.

Consider a system with an isotropic luminosity $dL_\nu$, at some
frequency $\nu$ (both in the fluid comoving frame), that is moving
with Lorentz factor $\gamma$ in a direction $\mu=\cos \theta$ with
width $d\mu$ relative to the line of sight.  Each photon of frequency
$\nu$ will be Lorentz boosted to frequency $\nu/\gamma(1-\beta\mu)$.
Using Lorentz transformation, the observed flux at distance $D$ from
the source satisfies:
\begin{equation}
dF_\nu=dL_{\nu \gamma (1-\beta \mu)} 
\frac 1 {4\pi D^2 \gamma^3(1-\beta\mu)^3}
\end{equation}

A spherical system, with a luminosity $L_\nu(R)$ in its local frame,
is simply a collection of such directed systems. The luminosity coming
from an angle with $d\mu$ around $\mu$ is $dL_\nu=d\mu L_\nu/2$ so that
\begin{equation}
dF_\nu=d \mu L_{\nu \gamma (1-\beta \mu)} 
\frac 1 {8\pi D^2 \gamma^3(1-\beta\mu)^3}.
\end{equation}
Since we are interested in the radiation coming from any $R_\perp$ per unit
area (surface brightness) this expression must be multiplied by 
\begin{equation}
\left| \frac {d\mu} {\pi dR_\perp^2} \right| = \frac 1 {2\pi R_\perp} 
\frac {d\mu} {dR} / \left| \frac {dR_\perp} {dR} \right|,
\end{equation}
where we can use equations \ref{muofR} and \ref{Rpofmu} 
to evaluate the derivatives. Using equation \ref{muofR} and assuming
$\beta\cong 1$, we can approximate
\begin{equation}
1-\beta\mu \cong 1-\beta+1-\mu \cong \frac 1 {16\gamma_l^2}
\left[ \left(\frac R {R_l} \right)^{-1} + 7 \left(\frac R {R_l} \right)^3 
\right]
\end{equation}
So that
\begin{equation}
\frac{dF_\nu}{\pi dR_\perp^2}=
\frac{256 \gamma_l^3 L_{\nu\gamma(1-\beta\mu)}}{\pi^2 D^2 R_l^2}
\frac{\left(\frac R {R_l}\right)^{11/2}
\left[1 +3\left(\frac R {R_l}\right)^4 \right]}
{\left[1-5\left(\frac R {R_l} \right)^4 \right]
\left[1+ 7\left(\frac R {R_l} \right)^4 \right]^3}
\end{equation}

We now assume that the luminosity in the local frame is changing with
radius and frequency as
\begin{equation}
\label{powerlawL}
L_\nu \propto R^{a}\nu^b.
\end{equation}

With this assumption the surface brightness coming from a radius $R$ is 
proportional to 
\begin{equation}
\label{sbofR}
\frac {(R/R_l)^{a-\frac 5 2 b+\frac {11} 2}} 
      {\left[ 1+ 7(R/R_l)^4 \right]^{3-b}}
\frac {1+3(R/R_l)^4} 
      {1-5(R/R_l)^4}.
\end{equation}
This is also a function of $R_\perp$ using equation \ref{Rpofmu}. Note
that the function $R_\perp(\mu)$ is double valued. For each $R_\perp$
there are two values of $\mu$, one from the front of the expanding
shell and one from its back. On the transition point, where
$R/R_l=0.67$, we get the relativistic version of limb brightening, as
the denominator of the second term in equation \ref{sbofR} vanishes.
The limbs in the relativistic version are not the edges of the surface
in a given time as in the nonrelativistic case but the edges of
surface from which photons arrive in a given observed time.
 
The assumption of equation \ref{powerlawL} is justified for frequencies
which are far from the typical synchrotron frequency (either above or
below). Fortunately these are two relevant cases as the radio range is
below the typical frequency for few months while the optical bands
are above the typical frequency after about 1 day. Assuming that the
electrons are not cooling significantly we find that the luminosity is
proportional to the number of electrons $\propto R^3$ times the power
produced by each electron $\propto \gamma^2 B^2\propto R^{-6}$.  This
is emitted at a typical frequency of $\nu_m\propto B \gamma^2\propto
R^{-9/2}$.  The luminosity at the typical frequency is therefor
\begin{equation}
L_{\nu_m}\propto R^{3/2}.
\end{equation}
The synchrotron low energy tail (assuming that the electrons are 
neither cooling nor self absorbed) is $\nu^{1/3}$ so that for $\nu\ll\nu_m$
\begin{equation}
L_\nu=L_{\nu_m}\left(\frac \nu {\nu_m}\right)^{1/3} 
\propto R^3\nu^{1/3} \ \ \rightarrow\ \  a=3,b=1/3.
\end{equation}
At the high energy tail, assuming that the electron distribution is a power
law with equal energy per decade we get
\begin{equation}
L_\nu=L_{\nu_m}\left(\frac \nu {\nu_m}\right)^{-1/2} 
\propto R^{-3/4}\nu^{-1/2} \ \ \rightarrow\ \  a=-3/4,b=-1/2.
\end{equation}

Figure 2 depicts the observed surface brightness for these two cases
of low and high frequencies. In both cases the surface brightness
diverges at $R_\perp=R_{\perp,\max}$. In the low frequency case the
surfaces brightness in the center is about half the average surface
brightness while in the high frequencies it is only about 10\%.
Practically all the luminosity at high frequencies is emitted from a
ring while at low frequencies it is spread more uniformly. This
qualitative difference might be intuitively understood as for high
frequencies the luminosity is decaying in time so that emission from a
lower radius (coming off the line of sight) is enhanced relative to
the emission from the line of sight.

\begin{figure}
\begin{center}
\plotone{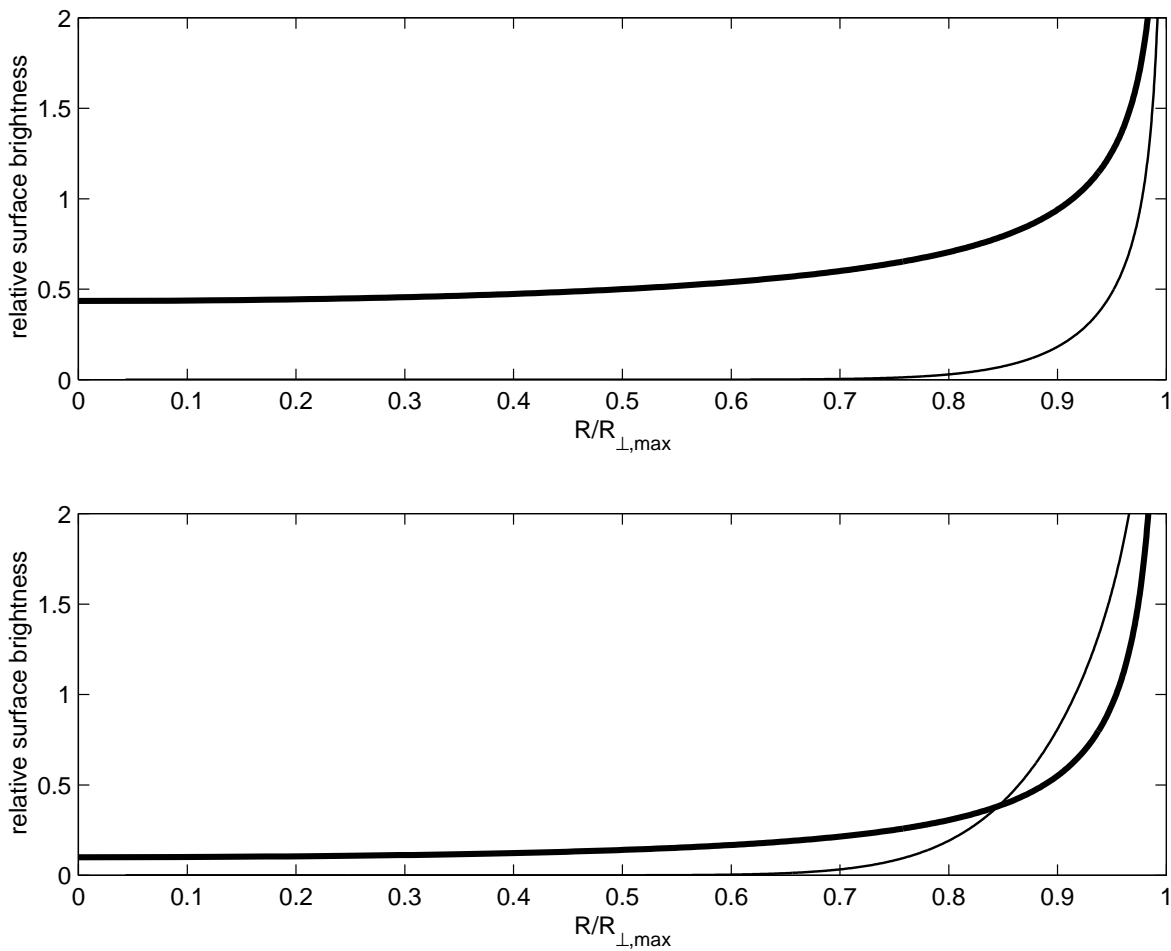}
\caption{Surface brightness as function of distance from the center.
  The upper frame shows the predicted surface brightness for
  frequencies below the typical synchrotron frequency and the lower
  frame is for frequencies above the typical synchrotron frequency.
  The heavy line is the contribution from the front of the shell while
  the light line is the contribution from the back. The surface
  brightness is normalized by the average surface brightness.}
\label{fracfig}
\end{center}
\end{figure}

As seen from figure 1, the radiation at a given observed time $T$
comes from a large range of radii $R$. However, using equation \ref{sbofR}
one can calculate the flux averaged radius of emission $\langle R
\rangle$ or the flux averaged perpendicular emission radius $\langle
R_\perp \rangle$. The Lorentz factor $\langle\gamma \rangle$ at the
average radius can then be calculated by $\langle\gamma
\rangle=(\langle R \rangle/R_l)^{-3/2}\gamma_l$. For low frequencies
we find
\begin{equation}
\label{ravlow}
\langle R \rangle=0.80 R_l \ \ , \ \ \langle R_\perp \rangle=0.81R_{\perp,max} \ \ , \ \
\langle \gamma \rangle=1.4 \gamma_l,
\end{equation}
while for high frequencies
\begin{equation}
\label{ravhigh}
\langle R \rangle=0.67 R_l \ \ , \ \ \langle R_\perp \rangle=0.90R_{\perp,max} \ \ , \ \
\langle \gamma \rangle=1.8 \gamma_l .
\end{equation}
Therefor radiation arriving the observer at a certain time in a
frequency band above the peak frequency, was emitted on average from
lower radius, higher Lorentz factor and outer $R_\perp$, than
radiation arriving to the observer at the same time in an energy band
below the peak. One can now find the relation between the typical
radius $\langle R \rangle$, the typical Lorentz factor $\langle\gamma
\rangle$ and the observed time $T$. For low frequencies:
\begin{equation}
T=\langle R \rangle/6.5 \langle \gamma \rangle^2c,
\end{equation}
while for high frequencies
\begin{equation}
T=\langle R \rangle/3.2\langle\gamma \rangle^2c.
\end{equation}
These replace the usual relation $T=R/2\gamma^2 c$. The coefficient in
the denominator depends on the frequency observed. Its range is
between $3.2$ (for frequencies far below the peak frequency) and $6.5$
(for frequencies far above the peak frequency). Therefor for the peak
frequency itself we expect the coefficient to be some average of
these two extreme values.

\section{Conclusions}
We have explored the spatial shape of the radiation emitted from a
decelerating relativistic shell in its adiabatic evolution $\gamma
\propto R^{-3/2}$. Using the self similar solution which gives the
exact proportionality constant we have arrived at an explicit
expression for the observed size of the emission as a function of
time.

We have further analyzed the observed surface brightness in two extreme
cases of frequencies far below and far above the typical synchrotron
frequency. These two extremes are of practical importance as the radio
is below the typical synchrotron frequency for few month and the
optical is above the typical synchrotron frequency after about a day.
We showed that this emission has sharp well defined edges. The edges
are brighter than the center as a result of a relativistic version of
the limb brightening effect.

The surface brightness for frequencies below the synchrotron typical
frequency was shown to be more uniform than in the high frequency
range in which most of the radiation comes from a small region near
the limbs. The exact shape is relevant for a detailed scintillation
theory. Furthermore, in the case of a ``nearby'' burst with
$z\cong0.1$ it is likely that the radio source will be resolved, and a
direct observation of the surface brightness will be possible.

The relation between the emission radius, the Lorentz factor and the
observed time is not well defined as there is a large range of radii
from which the emission reaches the observer at the same time. However
taking the flux averaged radius of emission and the Lorentz factor at
the average radius we find $T=\langle R \rangle/\alpha \langle \gamma
\rangle^2c$ where $\alpha=3.2$ for the high frequency limit and
$\alpha=6.5$ in the low frequency limit. These values of $\alpha$ are
smaller than the coefficient $\alpha=16$ obtained from the line of
sight (Sari 1997) however larger than the usual value of $\alpha=2$.

The effective Lorenz factor for high frequencies is larger by $1.3$
than the effective Lorentz factor for low frequencies. The transition
to the non relativistic stage and the effect of finite size jet
(Rhoads 1997) both depend on the Lorentz factor. Since the observed
time $T\propto\gamma^{-8/3}$, these effects would be observable in the
radio at about a factor of two earlier than they would be observed in
the optical range.

\acknowledgments

The author thanks The Center for Astrophysics for warm hospitality and
Tsvi Piran and Ramesh Narayan for helpful discussions. This work was
supported by NASA Grant NAG5-3516, and a US-Israel BSF Grant 95-328.

\end{document}